# Observation of 2nd band vortex solitons in 2D photonic lattices

Guy Bartal,[1] Ofer Manela,[1] Oren Cohen,[1] Jason W. Fleischer[2] and Mordechai Segev[1]

*1 Physics Department, Technion - Israel Institute of Technology, Haifa 32000, Israel*

*2 Department of Electrical Engineering, Princeton University, Princeton , New Jersey, 08544*

**Abstract**

We demonstrate second-band bright vortex-array solitons in photonic lattices. This constitutes the first experimental observation of higher-band solitons in any 2D periodic system. These solitons possess complex intensity and phase structures, yet they can be excited by a simple highly-localized vortex-ring beam. Finally, we show that the linear diffraction of such beams exhibits preferential transport along the lattice axes.

Wave propagation in nonlinear periodic structures exhibits many interesting phenomena [1]. In these settings, the dynamics is dominated by the interplay between the lattice properties and nonlinearity. A balance between the two may lead to the creation of self-localized waves: **lattice ("discrete") solitons** [1-6]. In particular, waves carrying angular momentum propagating in lattices of dimensionality two (or higher) exhibit a variety of fascinating phenomena related to the way coupling between lattice sites carries the angular momentum [7-9]. Experiments on such issues became feasible following the proposition [10] and the first demonstration [4] of optically-induced nonlinear photonic lattices. The first observation of 2D lattice solitons followed soon thereafter [11]. This has led to further exciting experiments in 2D nonlinear photonic lattices, such as the first observation of $1^{st}$-band vortex lattice solitons [12], 2D vector lattice solitons [13], 2D dipole-type lattice solitons [14], etc. Notwithstanding this important experimental progress in 2D lattice solitons, until recently, all research was limited to lattice solitons that originate from the $1^{st}$ band [11-15]. Last year, however, our group has predicted the existence of 2D lattice solitons arising from the X symmetry points in the transmission spectra of the $2^{nd}$ band [16]. These solitons possess the ring-shaped intensity profile, with a unique phase structure resembling a counter-rotating vortex array. Here, we present the first experimental observation of these $2^{nd}$-band lattice solitons, which reside in the gap between the $1^{st}$ and the $2^{nd}$ bands of the transmission spectra of a square lattice. ***This constitutes the first observation of any $2^{nd}$ band 2D soliton and the first experimental observation of any 2D gap soliton carrying angular momentum (vorticity).*** Surprisingly, our experiments reveal that, under proper nonlinear conditions, a simple vortex ring excitation naturally evolves into the $2^{nd}$-band vortex-ring lattice soliton, acquiring its unique counter-rotating vortex-array phase structure during propagation. At low-intensities, such $2^{nd}$-band excitations display intriguing features of preferential linear diffraction along the lattice axes, which stand in sharp contrast to the diffractive behavior of 1st-band vortex excitations which takes on the (square) symmetry of the lattice. We demonstrate these 2nd-band vortex-ring lattice solitons, and study their evolution dynamics experimentally and numerically. Such solitons can be observed in a variety of other systems, such as nonlinear fiber bundles, photonic crystal fibers, and Bose-Einstein condensates

We start by revisiting our recent paper [16] predicting 2nd-band vortex lattice solitons. Such solitons can be viewed as a superposition of two degenerate modes (same propagation constant $\beta$) of the defect they jointly induce (Fig. 1). One of these modes is associated with the 2nd-band X point ($k_x = \pi/D, k_y = 0$), and it is jointly trapped with a $\pi/2$ phase-delayed mode associated with the X' point ($k_x = 0, k_y = \pi/D$) of the same band [$D$ being the lattice period; see Fig.1b]. The combined mode displays a phase structure of a **2D array of vortices** with alternating rotation between neighboring sites (Fig. 1c) [17].

Our experiments are performed in a photorefractive crystal that possesses the (saturable) photorefractive screening nonlinearity [23]. The paraxial dynamics of a linearly polarized beam may be modeled by the non-dimensional equation [8]

$$i\frac{\partial \Psi}{\partial z} + \left(\frac{\partial^2 \Psi}{\partial x^2} + \frac{\partial^2 \Psi}{\partial y^2}\right) - \left[\frac{1}{1+V(x,y)+|\Psi|^2}\right]\Psi = 0 \qquad (1)$$

where $V = V_0\left[\cos(2\pi(x+y)/D) + \cos(2\pi(x-y)/D)\right]^2$ is the square lattice, $V_0$ the peak intensity of the lattice, $D$ the lattice spacing, and $\Psi$ is the slowly-varying amplitude of the electric field. The lattice potential is written in this fashion because in the experiment, the lattice is optically-induced in the photorefractive crystal by interfering four waves of ordinary polarization, while the probe (soliton-forming) beam is extraordinarily polarized along the crystalline c-axis [11,12]. Typical calculated results are shown in the upper row of Fig. 2. We propagate numerically a beam with the intensity and the phase of a vortex ring soliton (Fig. 2a,b, respectively) as calculated through the self-consistency method [19]. Under linear conditions, the input beam experiences diffraction in the square lattice, exiting after 5mm of propagation with the intensity structure shown in Fig. 2c. Notice the preferential transport along the lattice axes, typical to X-point excitations in momentum space. On the other hand, as we showed in [16], under the proper nonlinear conditions, the beam exhibits stable stationary propagation, exiting the lattice with an intensity (Fig. 2d) and phase (Fig. 2e) structure identical to that of the input.

Solitons, however, are not only stable, but are also robust. That is, an input wavepacket with parameters close enough to those of a soliton reshapes and naturally evolves into a soliton while shedding some of its power. Such robustness was demonstrated experimentally with spatial solitons in homogenous media [20], as well as with lattice solitons [4,11,12,21]. In fact, many lattice soliton experiments have used an input beam different than the soliton wavefunction, yet this input evolved into a soliton after some propagation distance in the lattice. More recently [21], it has been shown the power spectrum of a spatially incoherent input beam evolves into the characteristic power spectrum of a random-phase lattice soliton [22]. Here, we show (for the first time), that also the phase of a coherent input beam evolves and attains the soliton phase structure. Naturally, during this process some of the power radiates, while some transfers to the soliton mode. Yet, once the wavefunction has evolved into a soliton, it maintains its shape throughout propagation. The lower row in Fig. 2 shows the evolution of a simple singly-charged vortex-ring of a proper width (Figs. 2f, 2g). Under linear conditions, the ring broadens after 5 mm propagation (Fig. 2h). Under appropriate nonlinear conditions, the input evolves into a $2^{nd}$-band vortex lattice soliton, acquiring intensity and phase structures (Figs. 2i,2j) almost identical to those of the soliton (Figs. 2d,2e).

Our experiments are performed using a 488nm wavelength laser beam and a 5mm long SBN:75 photorefractive crystal displaying the screening nonlinearity, which is controlled by applying voltage along the crystalline c-axis [18]. Two pairs of plane waves interfere to optically-induce a square lattice inside the crystal, with each wave having a peak intensity of ~15mW/cm$^2$. In this case, the lattice spacing is 13µm, with each waveguide having a diameter of ~6µm. These plane waves are ordinarily-polarized hence they propagate linearly in the crystal, inducing a z-invariant photonic lattice [10,4]. The applied external field is chosen so as to construct a lattice with a maximum index modulation of 0.001 [6,11]. The probe beam is polarized extraordinarily, so that it experiences both the periodic refractive index, and the photorefractive screening nonlinearity, being able to form a soliton under proper parameters. We generate the input beam by reflecting the probe beam off a vortex mask of unity topological charge and then imaging the ring beam onto the crystal input face. The width of the input ring beam is comparable to the size of a single waveguide, and is launched around a single waveguide. Having a ring smaller than the lattice spacing corresponds to having its k-spectrum extending beyond the $1^{st}$ Brillouin zone, and launching the ring around a single waveguide maximizes the preferential excitation of modes from the $2^{nd}$ band.

Our experimental results are shown in Figs. 3-5. In Figure 3 we demonstrate the evolution of the simple vortex ring into a $2^{nd}$ band vortex soliton. Figure 3a shows the intensity pattern of the initial ring entering the lattice. Figure 3b shows the output (linear) diffraction pattern of a low-intensity beam after 5mm propagation in the lattice, exhibiting preferential diffraction along the lattice axes but retaining the size of the central ring. This feature of linear diffraction in the lattice stands in sharp contrast to that of a $1^{st}$-band vortex, which diffracts in a square formation (characteristic of the square lattice) [12], with both the central "hole" and the width of the ring expanding by the same amount. As we increase the probe beam intensity to one-half of the lattice peak intensity, the input beam reshapes and forms a $2^{nd}$-band vortex lattice soliton (Fig. 3c).

We visualize the phase structure of our beam by photographing the interference pattern it forms with a weakly-diverging Gaussian beam. Figure 4a depicts the phase of the simple vortex-ring beam entering the lattice, while Fig. 3b shows the phase of the beam exiting the lattice, which has propagated under the proper nonlinear conditions and has evolved into a $2^{nd}$ band vortex lattice soliton. Clearly, the simple singly-charged vortex structure of the input beam (Fig. 4a) has evolved into a more complex phase structure (Fig. 4b). We compare our experimental result of Fig. 3b to the phase structure of the calculated $2^{nd}$ band vortex lattice soliton by numerically "interfering" the calculated wavefunction with a weakly-diverging Gaussian beam. This interference pattern is shown in Fig. 4c, and is in excellent agreement with the experimental interference pattern of Fig. 4b (with no "fitting parameters"). This observation is also in a good agreement with our prediction shown in Fig. 2, where the simulations show that the simple vortex-ring acquires the soliton phase structure during propagation.

The unique diffraction behavior of the X-point $2^{nd}$ band excitations (Figs. 2c,h, 3b), along with the formation of the $2^{nd}$ band vortex lattice soliton evolving from a simple vortex ring, are intimately related to the phase evolution demonstrated in Fig. 4. In fact, these phenomena can be jointly clarified by examining the k-space distribution of the excitation wavepacket, and of the soliton arising from it. Figure 5a depicts the power spectrum of the excitation wavepacket (the simple vortex ring), with respect to the first Brillouin zone (BZ) [21,23]. The corners of the first BZ are the four bright dots which are the four beams constructing the lattice in momentum space [Footnote]. It is apparent that the input vortex ring excites Bloch modes primarily from higher bands, mostly from the vicinity of the X symmetry points of the $2^{nd}$ band. This excitation is responsible for the linear diffraction of the vortex ring in the lattice. The anisotropy of the band curvature at the vicinity of the four X symmetry points, together with the phase relation between these four points (determined by the vortex phase structure), result in the diffractive behavior presented in Fig.3, exhibiting preferential diffraction along the lattice axes. As stated above, this stands in a sharp contrast to the diffractive behavior of a $1^{st}$-band vortex, whose diffraction pattern takes on the (square) symmetry of the lattice [12], exemplifying the more isotropic band curvature at $\Gamma$ point in k-space.

As we increase the intensity of the probe beam to form the soliton, thereby increasing self-focusing, the modes from the anomalous diffraction region s transfer their power to the modes in the normally-diffracting regions, which in turn, become localized. The outcome is the reshaping of the power spectrum of the beam. This is clearly shown in Fig. 5b, which displays the power spectrum of the beam exiting the lattice, which lies mostly at the outer sides of the lines connecting the corners of the first BZ. The evolution of the phase of the input beam, from a simple vortex to the phase structure of the soliton, is a direct consequence of the reshaping the power spectrum undergoes. The experimental results of Figure 5 clearly show this reshaping. **Hence, the spatial components of the self-trapped vortex beam exiting the 2D lattice, after reshaping and forming a soliton, are those that belong to the normal diffraction regions of the $2^{nd}$ band, i.e., the X-points of the $2^{nd}$ band. Taken together, Figs. 3b, 3c, 4b and 5b show conclusively that we have experimentally demonstrated a $2^{nd}$-band vortex-ring lattice soliton.**

Finally, it is interesting to compare the power spectrum of the $2^{nd}$ band vortex soliton to the power spectrum of its $1^{st}$ band counterpart. Figure 5c displays the power spectrum of the $1^{st}$ band vortex soliton [12]. Clearly, the power spectrum of the $1^{st}$ band vortex soliton lies within the first Brilluoin zone, and has an almost circular symmetry. This power spectrum of a $1^{st}$-band vortex excitation results in a diffractive behavior in a square formation, with no preference along the axes, which is in a sharp contrast to the diffractive behavior of our $2^{nd}$-band vortex ring excitation, and to the phase structure of the $2^{nd}$-band vortex ring evolving from it.

In conclusion, we have presented the first experimental observation of a 2D higher-band soliton in a nonlinear periodic system, by demonstrating a second-band vortex soliton in a 2D photonic lattice. Such solitons have a complex phase structure resembling a counter-rotating vortex array that can be excited even with a simple single-vortex input beam. This nonlinear evolution has been demonstrated here both numerically and experimentally. As such higher-band vortex behavior is universal, we expect similar solitons to be observed in a variety of other systems (e.g. matter waves in optical lattices) in the near future.

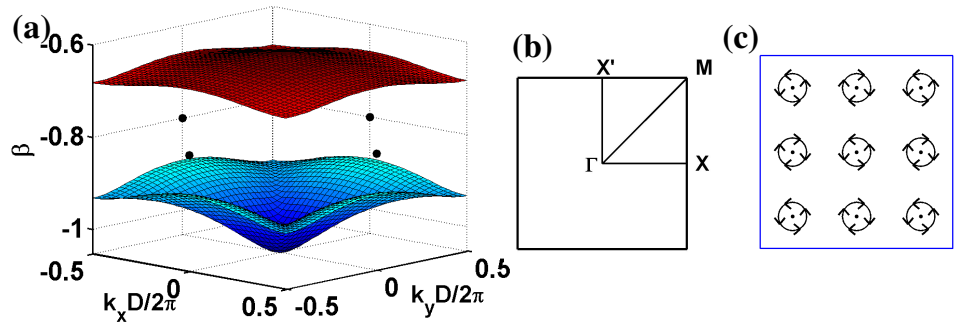

**Figure 1:** (a) The first two bands of the transmission spectrum in a 2D square lattice with D=10. The four thick dots mark the X-symmetry points. (b) High symmetry points of the reciprocal lattice. (c) Phase structure of a counter rotating vortex array, with the arrow in each vortex showing the direction of increasing phase.

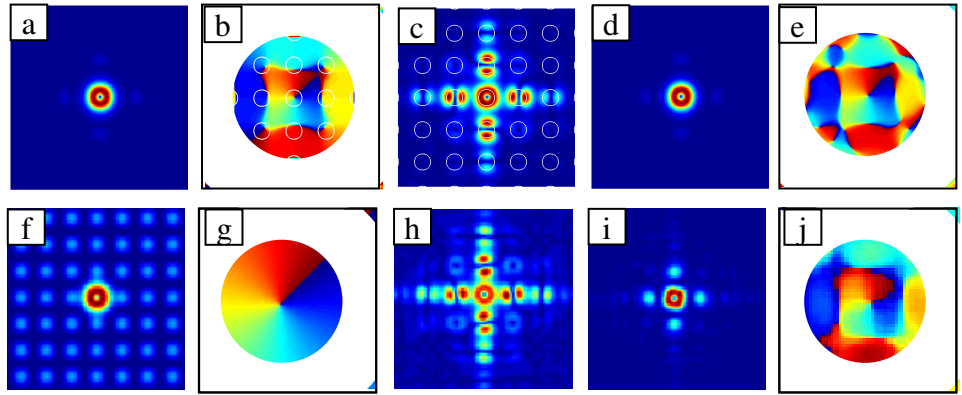

**Figure 2:** Simulated propagation of a $2^{nd}$-band vortex lattice soliton as calculated from self-consistency (upper row), and compared with a simulated evolution of a vortex ring with a simple (single vortex) phase structure into a $2^{nd}$-band vortex lattice soliton (lower row). **Upper row**: (a,b) Intensity and phase structure of the soliton as calculated via the self-consistency procedure, and used as the initial condition at the input plane of the lattice. (c) Linear diffraction (nonlinearity off) after 5mm of propagation. (d,e) Intensity and phase of the output beam under appropriate nonlinear conditions. (d,e) are identical to (a,b), thus confirming the beam is indeed a $2^{nd}$ band lattice soliton. **Lower row:** (f,g) Intensity (superimposed on the lattice) and phase structures of the input ring beam bearing a single vortex. (h) Linear diffraction (nonlinearity off) after 5mm of propagation. (i,j) Intensity and phase of the emerging output beam under appropriate nonlinear conditions. The simple ring excitation of (f,g) evolves into a structure (i,j) that is almost identical to the ideal structure of a the $2^{nd}$ band vortex lattice soliton.

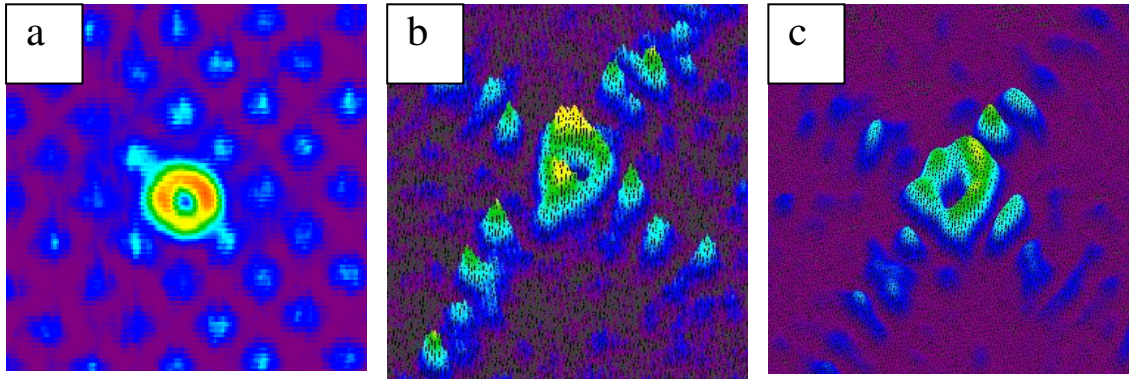

**Figure 3.** Experimental observation of a second-band vortex lattice soliton. (a) Intensity distribution of the input vortex-ring beam photographed (for size comparison) on the background of the optically induced lattice. (b) Output intensity distribution of a low intensity ring beam experiencing linear diffraction in the lattice. (c) Output intensity distribution of a high intensity ring beam, which has evolved into a second-band vortex soliton in the same lattice as in (b).

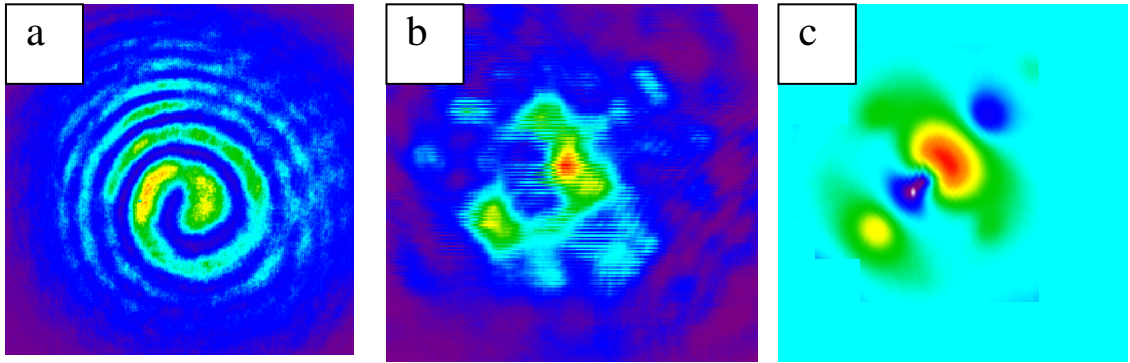

**Figure 4.** Experimental observation of the phase structure of a second-band vortex lattice soliton. The phase information is obtained by interference with a weakly diverging Gaussian beam. (a) Phase distribution of the input vortex-ring beam. (b) Output phase distribution of a high intensity ring-beam that has evolved into a second-band vortex soliton (interference pattern between the soliton whose intensity is presented in Fig. 3c and a weakly diverging Gaussian beam). (c) Numerical validation of the phase information in (b).

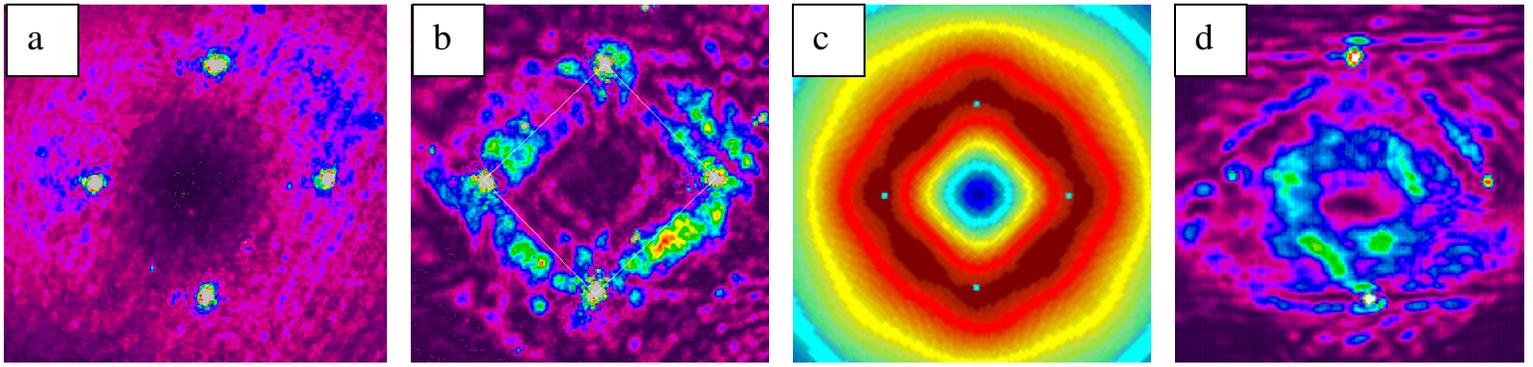

**Figure 5**. Power spectrum of the input (a) and output (b) beams. The formation of the $2^{nd}$ band vortex soliton reshapes the power spectrum. (c) Calculated power spectrum of the $2^{nd}$ band vortex soliton. (d) For comparison, power spectrum of the $1^{st}$ band vortex soliton of [15]. In all pictures, the "corners" of the $1^{st}$ BZ are the four lattice-forming beams, represented in the Fourier picture as four dots, marked with white circles.